# Microfacet projected area-based correction for *unified* model of Geant4 for rough surfaces

A. Morozov

LIP-Coimbra, Departamento de Física, Universidade de Coimbra, Rua Larga, 3004-516 Coimbra, Portugal

Email: Andrei.Morozov@coimbra.lip.pt

## Abstract

A modification of the optical model for rough surfaces, implemented in Geant4 as a part of the *unified* model, is suggested. The modified model takes into account the variation of the interaction probability of the photon with the microfacet based on the relative orientation of the photon and the sampled microfacet's normal. The implementation is using a rejection algorithm and assumes the interaction probability to be proportional to the projection of the microfacet area on the plane perpendicular to the photon direction. A comparison of the results obtained with the original and the modified models, as well as obtained in direct Monte Carlo simulations are presented for several test surfaces constructed using a pattern of elementary geometrical shapes.

## 1. Introduction

The *unified* model [1] implemented in the Geant4 toolkit [2-4] can be used to represent rough surfaces in optical simulations. It is assumed that the surface is made up of microfacets and that the angle between two normals, to the microfacet and to the "global" surface, follows a gaussian distribution. As the standard deviation is a user-defined parameter, the model enables simulation of surfaces with a broad range of roughness. A clear physical mechanism, generality and reliance only on one parameter which can be evaluated from experimental measurements [5] are the main advantages of the model, however, consideration of a standalone microfacet implies the absence of masking and shadowing effects (see e.g. [6]).

The *unified* model has another, more general, weakness related to the fact that the generated orientation of the microfacet is only weakly affected by the direction of the incoming photon. For every call of the model, triggered by the arrival of a photon at the surface, the only check which is made on the generated normal vector is preventing the microfacet from being oriented with its "back" toward the photon direction. The model is not taking into account that the probability for the photon to interact with a microfacet of a given orientation depends on that orientation. More specifically, the probability should be proportional to the projection of the microfacet area at the plane perpendicular to the direction of the photon ("visible" area of the microfacet), which, in more advanced models is typically described as a part of the so called shadowing-masking function (see e.g. [7]). This paper demonstrates how implementation of a simple rejection-based mechanism can improve the predictive power of the *unified* model.

# 2. Methods

## 2.1 Microfacet's normal generation: current implementation in Geant4

In the Geant4 version 11.3.0 (the newest at the moment of writing this paper), the generator of the microfacet normal vector is implemented in the GetFacetNormal method of the G4OpBoundaryProcess class [8]. The algorithm is the following:

1. Sample the angle alpha between the normals to the microfacet and to the global surface using a rejection based random generator
2. Sample the angle phi, giving the rotation of the microfacet normal around the global one (uniform in the range from 0 to 2π)
3. Construct the microfacet's normal unit vector
4. Compute the scalar product of the photon direction vector and the microfacet's normal
5. Reject the normal (and repeat the generation cycle) if the scalar product is zero or positive

The check made in step (5) rejects the normals which result in the photon arriving at the microfacet from the back or along the surface. Note that the positive normal direction in Geant4 is backward from the interface.

## 2.2 Microfacet normal generation: suggested modification

The current implementation, described in the previous chapter, does not take into account the fact that the normals giving larger projection of the microfacet's area at the plane perpendicular to the photon direction ("projected" or "visible" area) should have higher generation probability. The most straightforward modification is to keep the steps from 1 to 4 of the current implementation and only modify the rejection mechanism. The following criteria should be considered:

a) Always accept the normal if the photon is normally incident at the front surface of the microfacet (local incidence angle of 0 degrees)
b) The probability to keep the generated normal should decrease with the reduction of the projected area, reaching 0 at 90 degrees of the local angle of incidence
c) As in the current implementation, forbid local incidence angles above 90 degrees

The scalar product of the photon direction and the normal gives the cosine of the local incidence angle at the microfacet with the negative sign. The visible area of a microfacet with unitary area is equal to that cosine. Therefore, the acceptance of the normal can be formulated as the negative of the scalar product is larger than a random number uniformly generated in the range from 0 to 1. Equivalently, the rejection criterion is that the scalar product is larger or equal to the generated random number.

## 2.3 Validation technique

A rough surface was modelled by placing a repeating pattern of geometric shapes (cones or half-spheres) on top of a plane, as shown by an example in Figure 1. While not offering a realistic scenario, the capability to predict the distribution of the microfacet normals for the photons of a given direction, uniformly irradiating the surface, is still expected of the model. The advantage is that in this case the distribution of the angle between the global and the microfacet normal is exactly known.

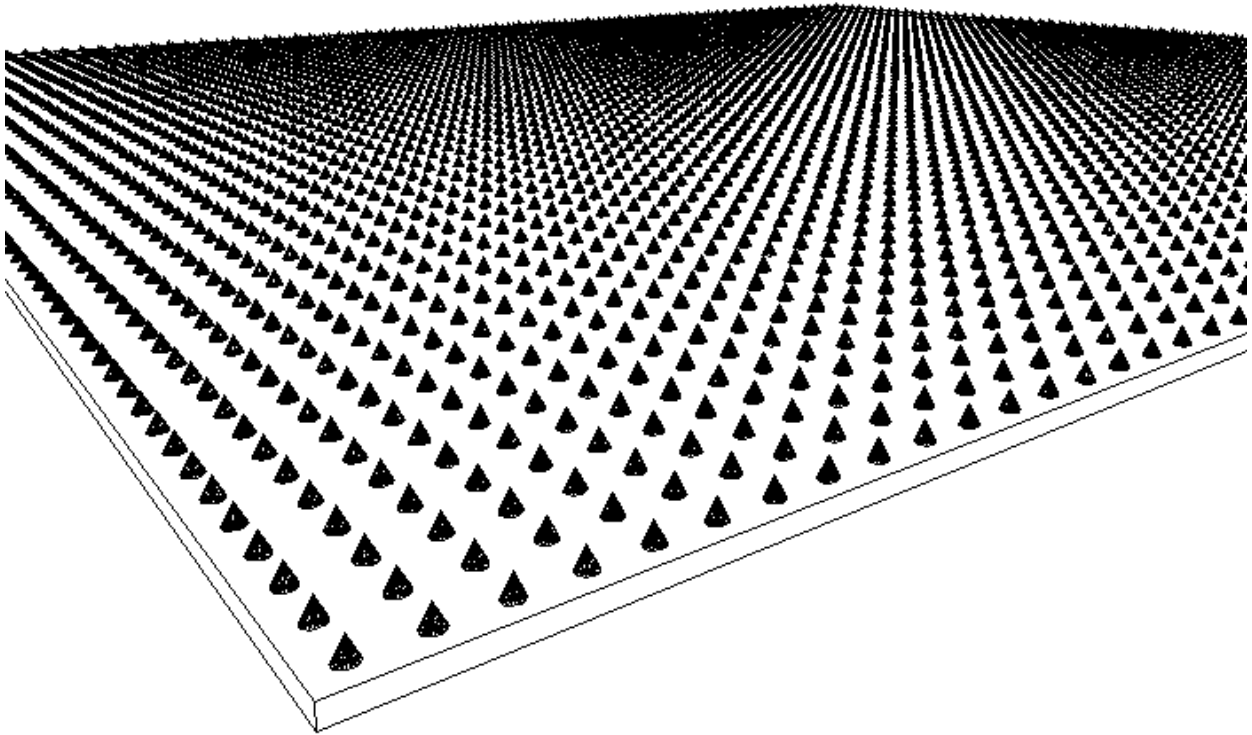

Figure 1. Example of the representation of a rough surface used in this study: a regular pattern of cones on top of a plane.

$10^6$ photons were uniformly generated in a square area above the surface. The size of the area was selected to be large in comparison with the pattern pitch and all photons were assigned with the same direction. Each photon was traced until intersecting with the surface (plane or one of the shapes) and the angle between the local normal and the photon direction, known in the simulation, was stored.

On the other hand, the generators of the microfacet normal vector were constructed following the approaches described in sections 2.1 and 2.2. The only modification was introduced in step (1) to take into account the true distribution of the angle between the microfacet's and the global normals. For the case of the cones, there are only two angles: one for the flat surface (0 degrees) and the other one for the cone's side surface. The statistical weights for these two values are given by the area ratio of the plane surface between the cones and the side surfaces of the cones. For the case of the half-spheres the random generator was implemented using the GetRandom method of the TH1 histogram of the ROOT toolkit [9]. The histogram was filled by sweeping the half-sphere area and extracting the angle between the normals. The band area was used as the statistical weight. The contribution from the flat surface was added to the histogram considering the area ratio of the flat and the half-sphere surfaces. Both generators (original and the area-corrected methods) were then run $10^6$ times using a given photon direction. For each generated normal the angle between the local normal direction and that of the photon was computed and stored.

Finally, the distributions of the angle between the local normal and the photon direction, obtained with the Monte Carlo method (true data) and given by the generators, were compared.

## 3. Results and discussion

### 3.1 Validation using test surfaces

The output of the normal vector generators (with and without the area correction) were compared with the results of the Monte Carlo simulation using the approach described in the previous section. The distributions of the angle between the photon direction and the normal to the microfacet (local incidence angle) were first compared for the cones with the height to the base diameter ratio of 0.1 (representing a surface with low roughness) and the placement pitch of 3 diameters. Four global angles of incidence (angles in respect to the global normal) were tested: 0, 21.6, 45 and 71.6 degrees. Note that even for the largest angle there is no shadowing of the cones by each other. The obtained distributions are shown in Figure 2. Note that the modified method, in contrast to the original one, gives distributions very similar to the ones obtained in Monte Carlo simulations for all considered angles of incidence.

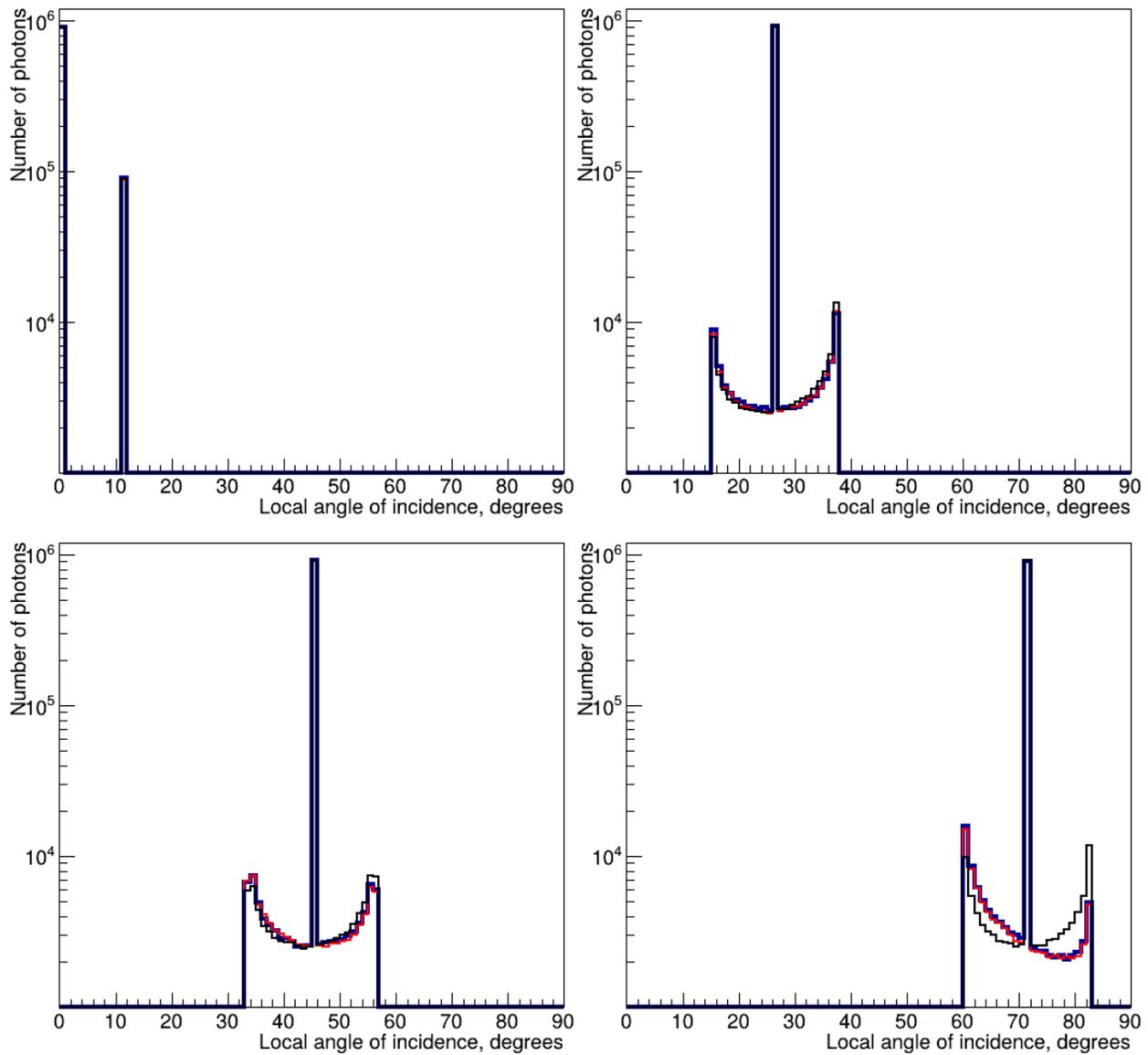

Figure 2. Distributions of the local angle of incidence for the case of the surface with cones of 0.1 height-to-diameter ratio: true data from the Monte Carlo simulations (thick blue line), prediction of the original (thin black line) and the modified *unified* model (thick red line). The global angle of incidence is 0 (top left), 21.6 (top right), 45 (bottom left) and 71.6 degrees (bottom right).

A surface with higher roughness was simulated by increasing the height of the cones to be equal to the base diameter. The results are shown in Figure 3, demonstrating that the area-corrected method gives significantly better predictions.

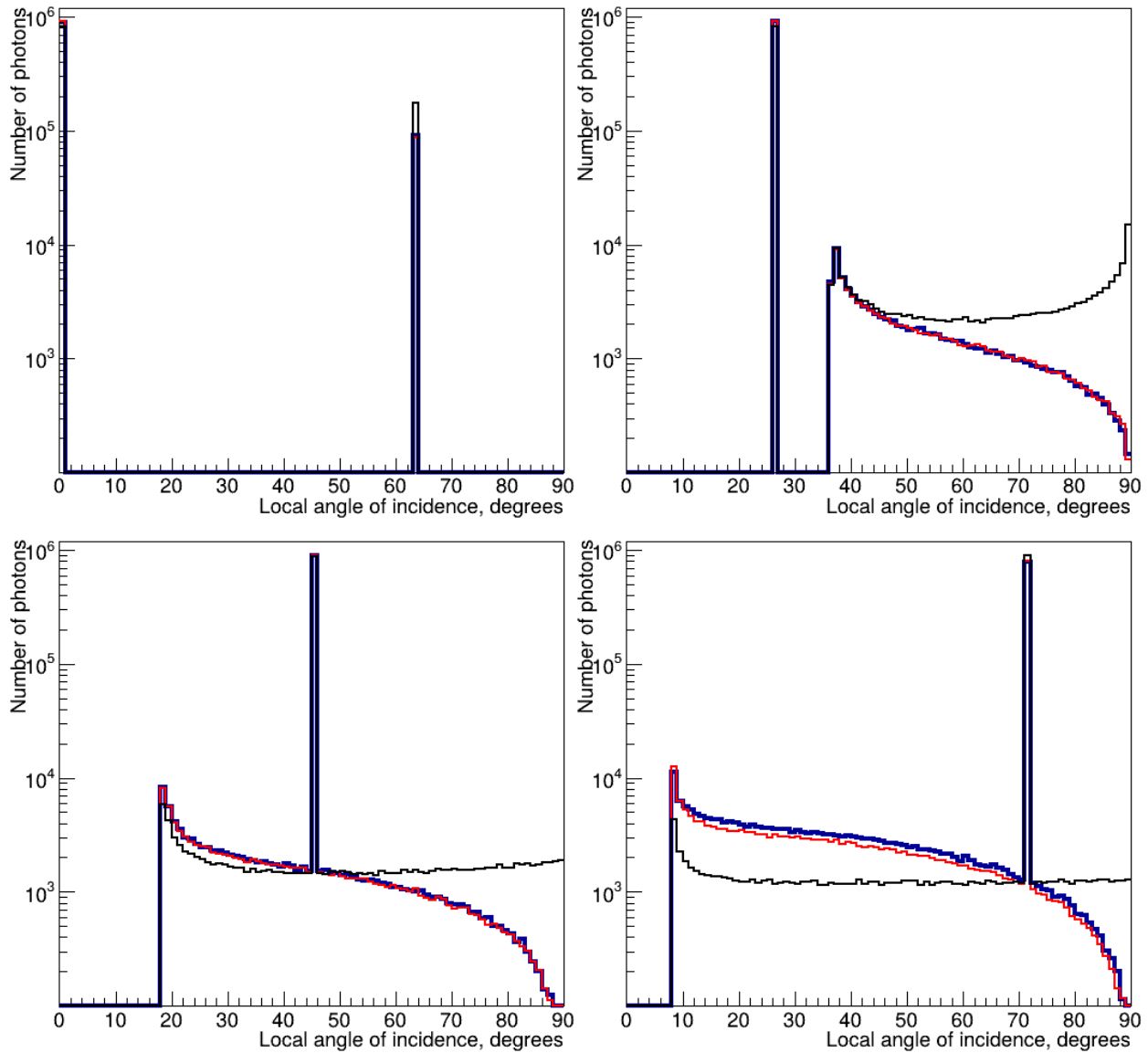

Figure 3. Distributions of the local angle of incidence for the case of the surface with cones of equal height and diameter: true data from the Monte Carlo simulations (thick blue line), prediction of the original (thin black line) and the modified *unified* model (thick red line). The global angle of incidence is 0 (top left), 21.6 (top right), 45 (bottom left) and 71.6 degrees (bottom right).

Another test surface was constructed with a pattern of half-spheres, positioned with the pitch of three diameters. The results for the same four global angles of incidence are shown in Figure 4. The corrected model also shows significantly better predictive power, especially at large angles of incidence (both local and global).

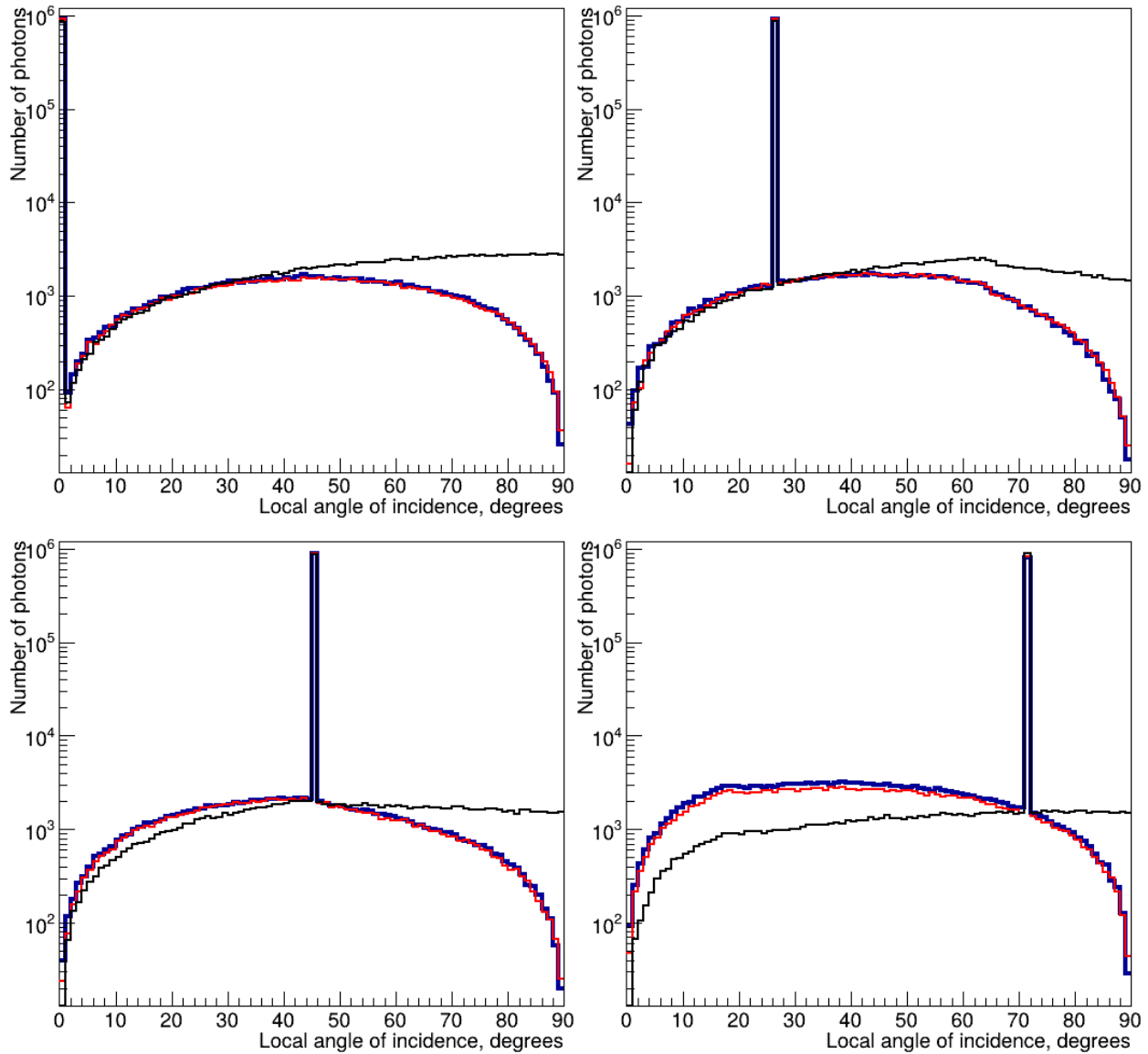

Figure 4. Distributions of the local angle of incidence for the case of the surface with half-spheres and the pitch of positioning equal to three diameters: true data from the Monte Carlo simulations (thick blue line), prediction of the original (thin black line) and the modified *unified* model (thick red line). The global angle of incidence is 0 (top left), 21.6 (top right), 45 (bottom left) and 71.6 degrees (bottom right).

As discussed above, the *unified* surface model does not take into account shadowing effects. The importance of shadowing can be demonstrated by reducing the pitch in the half-sphere positioning to one diameter. Note that the photons were incident at the surface along the direction in which half-spheres touch at their base. As shown in Figure 5, the shadowing effect is noticeable at 45 degrees and strong at 71.6 degrees. Considering the large divergence of the prediction of the original model from the Monte Carlo results at large local angles of incidence, the use of the area corrected method is still advantageous.

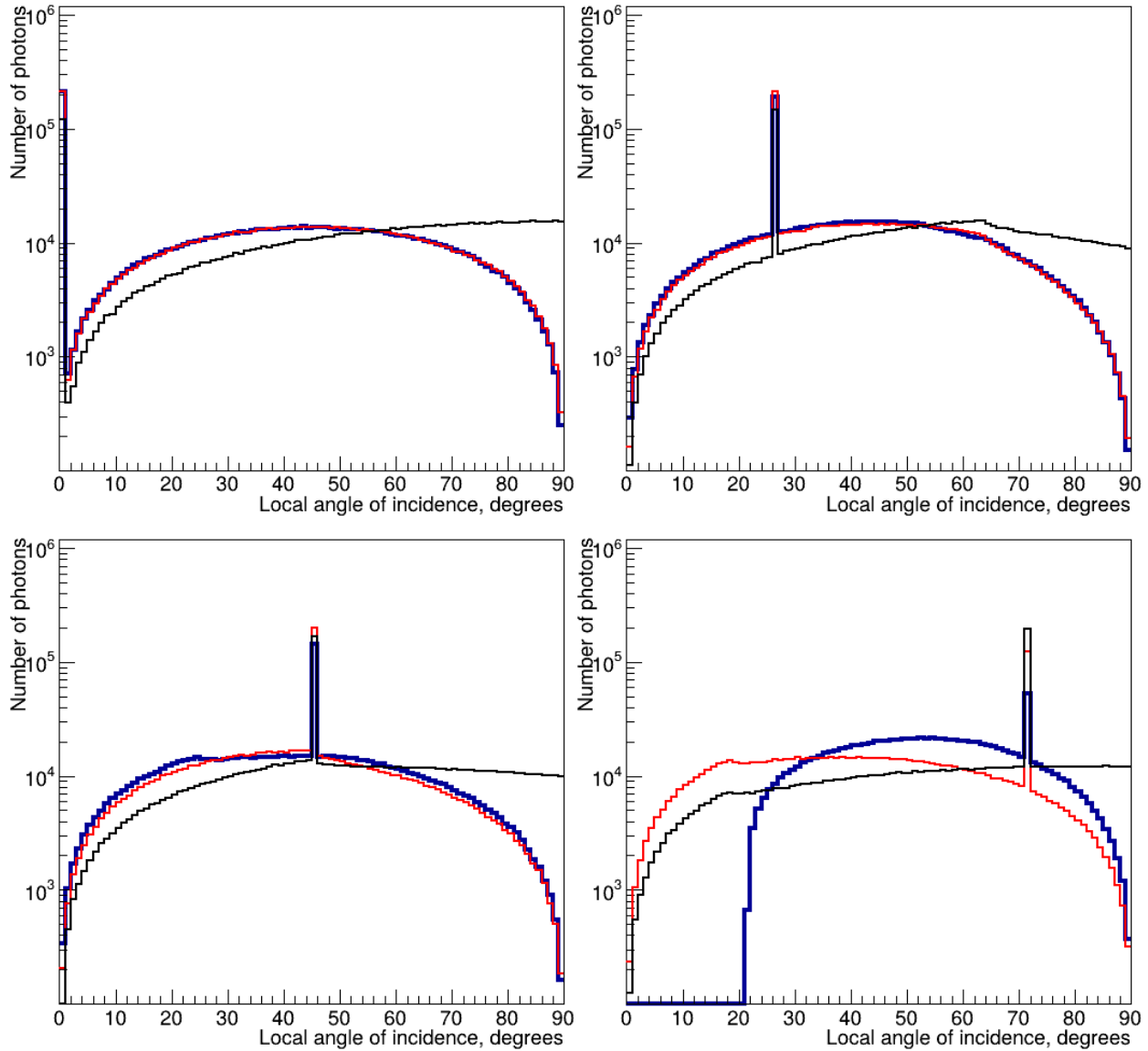

Figure 5. Distributions of the local angle of incidence for the case of the surface with half-spheres and the pitch of positioning equal to the diameter: true data from the Monte Carlo simulations (thick blue line), prediction of the original (thin black line) and the modified *unified* model (thick red line). The global angle of incidence is 0 (top left), 21.6 (top right), 45 (bottom left) and 71.6 degrees (bottom right). The strong deviation of the model predictions from the Monte Carlo data is caused by the shadowing of half-spheres by their neighbours.

## 3.2 Impact on photon transport

This section discusses the impact of the proposed modification of the model on the photon transport. The following tests were conducted using the standard assumption of the *unified* model: gaussian distribution of the angle between the microfacet's and the global normals. Two cases were considered: the standard deviations (*sigma alpha*) of 0.1 and 0.9 radians to represent a low and high roughness surface. The refractive indexes of the materials before and after the surface were chosen to be 1.81 and 1, which corresponds to the LYSO-to-air interface at the emission peak of the LYSO scintillator.

In the first test the rough surface was irradiated with $10^6$ photons with the global incidence angle of 71.6 degrees. The Fresnel equations were used to compute the unpolarized reflection coefficients. For the reflected photons the distribution of the global angle of reflection (measured from the global normal) was collected. The results for the original and the modified models are shown in Figure 6, the data for the low surface roughness on the left and for the high roughness on the right hand side. The difference between the results of the two models is quite significant.

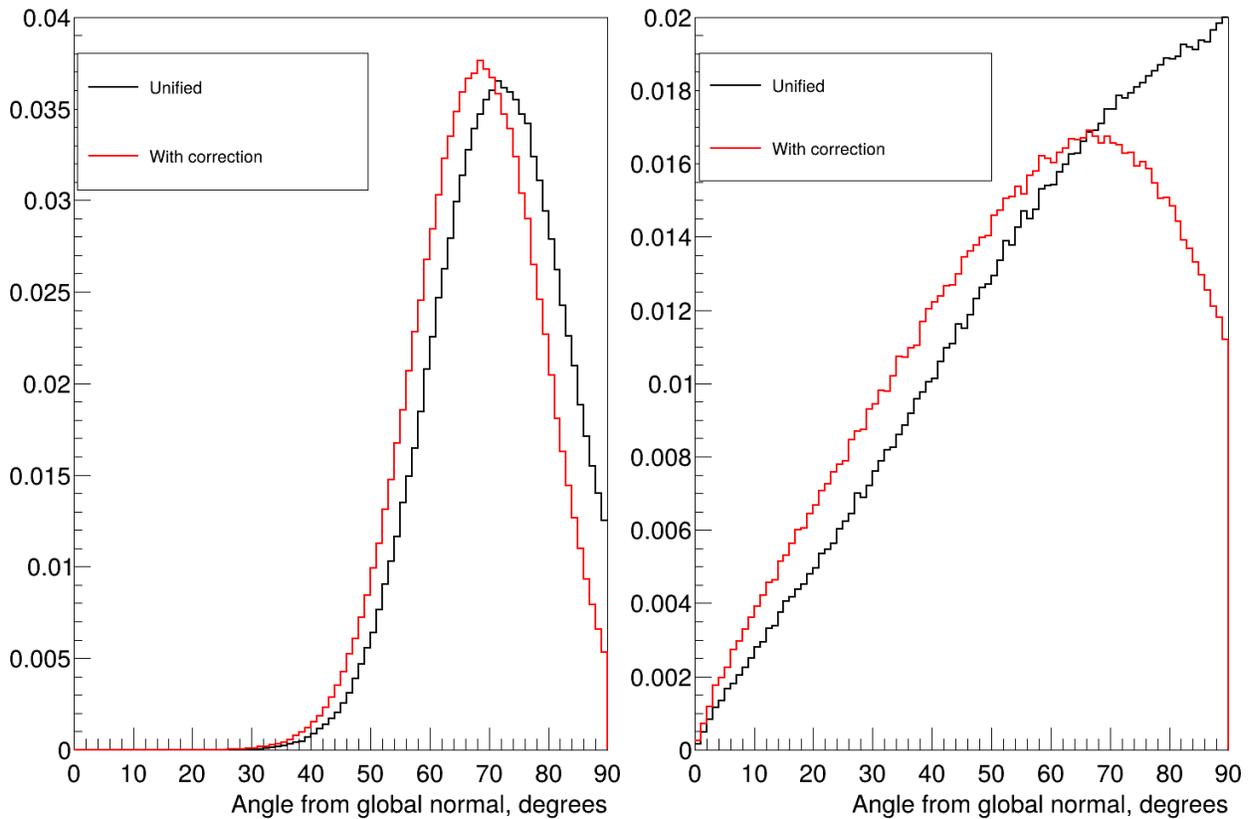

Figure 6. Distribution of the angle of reflection from the global normal for the original (thin black line) and the modified *unified* model (thick red line). The data are shown for the LYSO-to-air interface and the global angle of incidence of 71.6 degrees. The sigma alpha is 0.1 and 0.9 radians for the left and right hand side, respectively.

Another test was conducted to evaluate the fraction of photons, reflected from the same LYSO-to-air interface, this time as a function of the global angle of incidence. Both models give essentially the same result for the case of sigma alpha of 0.1 radians, while the predictions for the case with sigma alpha of 0.9 radians are significantly different as demonstrated in Figure 7.

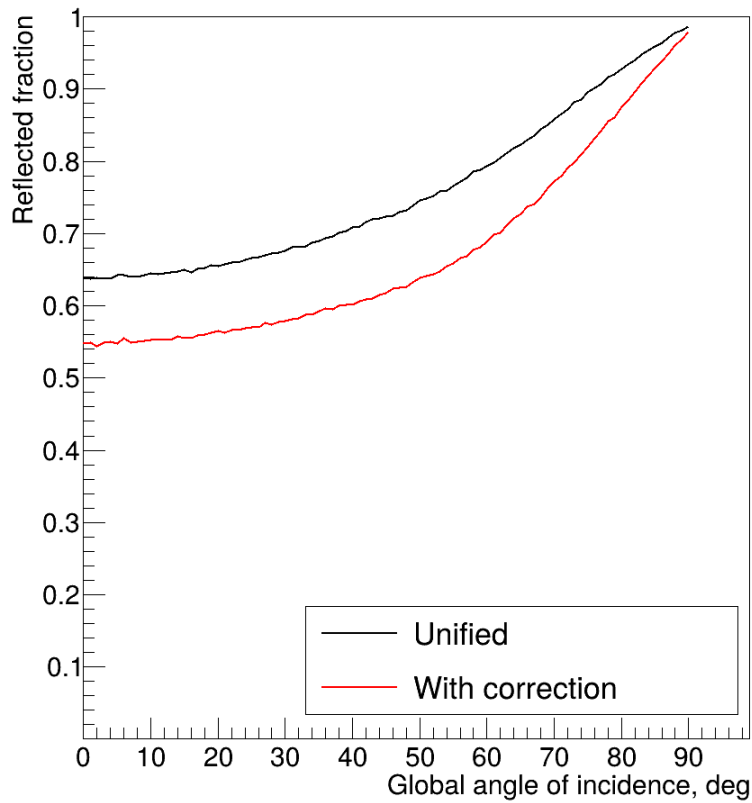

Figure 7. Fraction of the reflected photons from the LYSO-to-air interface as a function of the global angle of incidence: the original (thin black line) and the modified *unified* model (thick red line). The sigma alpha of the surface is 0.9 radians.

## 4. Conclusions

An improvement for the *unified* model, implemented in Geant4 for optical simulations of rough optical surfaces, is suggested. The modification adds a rejection-based mechanism to take into account the interaction probability of the photon with a microfacet based on the "visible" area of the microfacet. A comparison of the results obtained with the original and the modified models, as well as obtained in direct Monte Carlo simulations shows significantly improved predictive power of the modified model. Note that the *unified* model (both original and the modified version) do not consider the masking and shadowing effects as interaction of the photon is limited to a single microfacet, and use of a more advanced model is recommended for surfaces with high roughness.